\documentstyle[pre,aps,psfig]{revtex}
\draft

\begin{document}                

\def\be{\begin{equation}}
\def\ee{\end{equation}}
\def\ba{\begin{eqnarray}}
\def\ea{\end{eqnarray}}

\preprint{DOE/ER/40561-50}

\title{Lyapunov exponents and Kolmogorov-Sinai entropy for a 
high-dimensional convex billiard}
\author{Thomas Papenbrock}
\address{Institute for Nuclear Theory, Department of Physics, 
University of Washington, Seattle, WA 98195, USA}
\maketitle
\begin{abstract}
We compute the Lyapunov exponents and the Kolmogorov-Sinai (KS) entropy for a
self-bound $N$-body system that is realized as a convex billiard. This system 
exhibits truly high-dimensional chaos, and $2N-4$ Lyapunov exponents are found 
to be positive. The KS entropy increases linearly with the numbers of 
particles. We examine the chaos generating defocusing mechanism and 
investigate how high-dimensional chaos develops in this system with no 
dispersing elements.
\end{abstract}
\section{Introduction}

Billiards are simple yet nontrivial examples of systems that display
classically chaotic motion. Of special importance are the Sinai billiard
\cite{SinaiB} and the Bunimovich stadium \cite{BunimS} since they are known to
be completely chaotic.  Interestingly, these two systems exhibit two different
mechanisms that generate chaos. While dispersion is the chaos-generating
mechanism in the Sinai billiard it is defocusing that leads to chaotic dynamics
in the Bunimovich stadium. Dispersing yields a permanent divergence of
neighbored trajectories. Defocusing may occur upon reflection at a focusing
boundary element. Provided the free path is sufficiently long nearby
trajectories start to diverge after passing through the focusing point, and on
average the divergence might exceed the convergence thus leading to exponential
instability. Dispersing billiards are well known also in higher
dimensions. Popular examples are the three-dimensional Sinai billiard and the
hard sphere gas. However, it was not until recently that completely chaotic
billiards were constructed in more than two spatial dimensions that rely
entirely on the defocusing mechanism \cite{RehaBu96,BCG96}. These billiards 
use spherical caps as the focusing elements of the boundary. A trajectory
diverges mainly in a two-dimensional plane that is defined by the points of
consecutive reflections with the spherical cap, and focusing may be very weak
in the transversal directions. This makes it more difficult to create truly
high-dimensional chaos in focusing billiards than in dispersing ones. 
Sufficient conditions for the construction of high-dimensional focusing 
billiards were given in ref.\cite{RehaBu96}, but it was found that these are
not necessary ones\cite{BCG96}.

Besides of their intrinsic interest billiards are important
model systems in the field of quantum chaos \cite{GMW} and statistical and
fluid mechanics \cite{Gaspard}. Questions related to chaos, ergodicity, 
transport and equilibration are often studied in billiard models, see e.g. 
refs.\cite{DellPoschHoov96,vZvBD}. While two-dymensional chaos is fairly well 
understood by now, much less is known in high-dimensional systems. 
Recently, a high-dimensional billiard model has been proposed in the
context of nuclear physics \cite{TP99} and quantum chaos \cite{PP99}. Within
this model a self-bound $N$-body system is realized as a convex billiard.
Numerical computations yielded a positive largest Lyapunov exponent and showed
that this system is predominantly chaotic. In this article we want to extend
previous calculations and compute the full Lyapunov spectrum and the KS entropy
for this chaotic $N$-body system. These quantities characterize the degree of
hyperbolic instability in dynamical systems and may be related to transport
coefficients in non-equilibrium situations \cite{Ruelle}. Since the studied
billiard is convex, defocusing is the only possible source causing this
instability \cite{Bunim91}. This makes it interesting to examine this mechanism
in more detail and compare to the situation of defocusing billiards with
spherical caps. The results of such an investigation are not only of
theoretical interest but may also be useful for further applications. We have
in mind general questions concerning chaos in self-bound many-body systems like
nuclei or atomic clusters and its influence on equilibration, damping or
transport processes.

This paper is organized as follows. In the next section we describe the model
system and the techniques used to compute the Lyapunov exponents.  The third
section contains the results of our numerical computations for various system
sizes $N$. In section four we investigate the defocusing mechanism in more
detail. We finally give a summary.

\section{High-dimensional billiard and Lyapunov exponents}

Let us consider a classical system of $N$ particles with Hamiltonian
\be
\label{ham}
H=\sum_{i=1}^N{p_i^2\over 2m} + \sum_{i<j}V(|\vec{r}_i-\vec{r}_j|),
\ee
where $\vec{r}_i$ is a two-dimensional position vector of the $i$-th particle
and $\vec{p}_i$ is its conjugate momentum. The interaction is given by
\ba
\label{int}
V(r)=\left\{
     \begin{array}{ll}
     0 & \mbox{for $r<a$}, \\
     \infty & \mbox{for $r\ge a$}.
     \end{array}
     \right.
\ea
Thus, the particles move freely and interact whenever the distance between a
pair of particles reaches its maximum value $a$. Hamiltonian (\ref{ham})
defines a self-bound, interacting many-body system. Energy, total momentum and
total angular momentum are conserved quantities.  For large numbers of
particles the points of interactions are close to a circle of diameter
$a$ and therefore define a rather thin surface. Therefore, this system is 
a simple classical model for nuclei or atomic clusters.  For finite values 
of the binding potential the system is amenable to a mean field description
\cite{BPR99}. Hamiltonian (\ref{ham}) may also be viewed as a special case of
the square well gas \cite{SWG} with infinite binding potential. However, to the
best of our knowledge, the square well gas has not been investigated for such
parameter values. In what follows we restrict ourselves to the case of
vanishing total momentum and angular momentum. 

In the limit $N\to\infty$ the number density diverges for the self-bound
many-body system (\ref{ham},\ref{int}). A constant density may be obtained once
the parameter $a$ is rescaled as $a\to a N^{1/3}$, thus turning the Hamiltonian
(\ref{ham},\ref{int}) into an effective Hamiltonian. In what follows we work
with a $N$-independent parameter $a$. Since the billiard is a scaling system
one may easily rescale the results obtained below to adapt for different values
of $a$.

The time evolution of a many-body system with billiard like interactions
requires an effort ${\cal{O}}(N\ln{N})$ to be compared with the effort
${\cal{O}}(N^2)$ for a generic two-body interaction \cite{Prosen}. Initially
one computes the $N(N-1)/2$ times at which pairs of particles may interact and
organizes these in a partially ordered binary tree, keeping the shortest time
at its root. Immediately after an interaction of particles labelled $i$ and $j$
one has to recompute $2N-3$ times corresponding to future interactions between
particles $i$ and $j$ and the remaining ones. The insertion of each new time
into the partially ordered tree requires only an effort
${\cal{O}}(\ln{N})$. Between consecutive interactions particles move
freely. Upon an interaction of particles labelled by $i$ and $j$, respectively
the momenta change accordingly to 
\be
\label{reflect}
\vec{p}'= \vec{p} - 2{\vec{p}\cdot\vec{r}\over a^2}\vec{r}.
\ee  
Here, $\vec{p}'$ and $\vec{p}\equiv\vec{p}_i-\vec{p}_j$ are the relative
momentum vectors immediately before and after the interaction, respectively,
and $\vec{r}\equiv\vec{r}_i-\vec{r}_j$ is the relative position vector with
magnitude $|\vec{r}|=a$ at the interaction. Obviously, eq.~(\ref{reflect})
describes a reflection in the center of mass system of the two
interacting particles.

We now turn to the computation of the Lyapunov exponents. We describe the used
techniques rather briefly since a large body of literature exists on the
subject, see e.g. refs.\cite{Gaspard,LichtLieb,Reichl}.  A Hamiltonian system
with $f$ degrees of freedom possesses $f$ independent Lyapunov exponents
$\lambda_1,\ldots,\lambda_f$ ordered such that
$0\le\lambda_1\le\ldots\le\lambda_f$. Since the Hamiltonian flow preserves
phase space volume there are also $f$ non-positive Lyapunov exponents with
$\lambda_{-j}=-\lambda_j$. A system with $n$ integrals of motion has $n$
vanishing Lyapunov exponents $\lambda_1=\ldots=\lambda_n=0$, while a chaotic
system has a positive largest Lyapunov exponent $\lambda_f>0$.  This exponent
$\lambda_f$ is the rate at which neighbored trajectories diverge under the time
evolution.

Benettin {\it et al.} \cite{Benettin76} gave a method to compute the largest
Lyapunov exponent from following the time evolution of a reference trajectory
and a second one that is initially slightly displaced. The displacement vector
has to be rescaled after some finite evolution in a compact phase space.  To
compute the full spectrum of Lyapunov exponents one has to follow $f$
trajectories besides the reference trajectory \cite{Benettin79}. This defines
$f$ independent displacement vectors, and finite numerical precision requires
their reorthogonalization besides the rescaling during the time evolution.

Rather than following the time evolution of finite displacement vectors one may
also use infinitesimal displacements (tangent vectors) in the computation of
the Lyapunov exponents. In tangent space the time evolution is given by a
linear mapping. Details about the tangent map in high-dimensional billiards 
can be found in refs. \cite{DellPoschHoov96,Sieber}.

In a completely chaotic system the KS entropy is given by the sum of all
positive Lyapunov exponents \cite{Piesin76}, i.e. $h_{\rm KS}=\sum_{j=1}^f
\lambda_j$.  The KS entropy measures at which rate information about the
initial state of a system is lost. 

\section{Results}

In what follows we consider the $N$-body system at vanishing total momentum and
angular momentum. We use units such that $a=m=E/N=1$. Times are then given in
units of $a(mN/E)^{1/2}$. We choose initial conditions at random and follow a
trajectory for at least $10^6$ collisions. This ensures a good convergence of
the numerically computed Lyapunov spectra. 

We have checked our results as follows: The time evolution was checked by
comparing forward with backward propagation; the Lyapunov spectra were checked
by comparing the results obtained from the tangent map with those obtained by
Benettin's method involving finite displacements; the computation of all
Lyapunov exponents showed that $\lambda_{-j}+\lambda_j$ vanishes within our
numerical accuracy; we found four pairs of vanishing Lyapunov exponents
corresponding to the conserved quantities.

\begin{figure}
  \begin{center}
    \leavevmode
    \parbox{0.9\textwidth}
           {\psfig{file=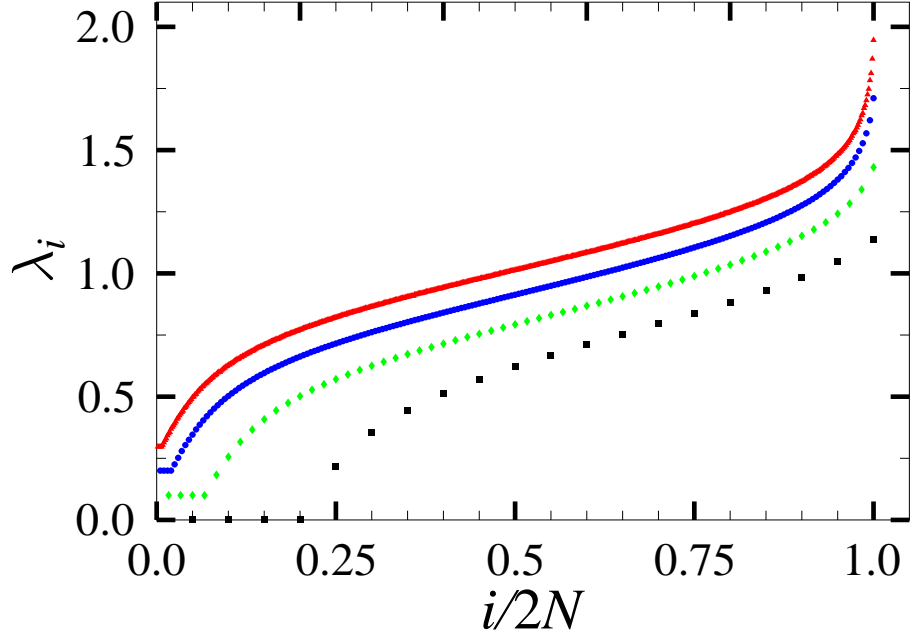,width=0.8\textwidth,angle=0}}
  \end{center}
\protect\caption{Lyapunov spectra for $N=10$ (squares), $N=30$ (diamonds),
$N=100$ (circles) and $N=300$ (triangles) in units of $(E/Nma^2)^{1/2}$.
The last three spectra are shifted by 0.1, 0.2 and 0.3$(E/Nma^2)^{1/2}$,
respectively.}
\label{fig1}
\end{figure}

The Lyapunov spectra for systems of sizes $N=10,30,100,300$ particles are
plotted in Fig.~\ref{fig1}. We note that the $N$-body system possesses $2N-4$
positive Lyapunov exponents.  This shows that there are no further integrals of
motion besides energy, momentum and angular momentum, and that truly
high-dimensional chaos is developed. We discuss this finding in detail 
in the following section.  The Lyapunov exponent $\lambda_i$ is a smooth
function of its index with a rather small smallest positive Lyapunov exponent
$\lambda_5$.  This behavior is similar to the case of the Lennard-Jones
fluid \cite{LJ} or the Fermi-Pasta-Ulam model \cite{FPU} but differs from the
hard sphere gas where a rather large smallest positive Lyapunov exponent was
found \cite{DellPoschHoov96}. Note that the spectra seem to converge somehow 
with increasing $N$. Table~\ref{tab1} displays the largest and smallest 
positive Lyapunov exponents, collision rates and the KS entropies.

\begin{figure}
  \begin{center}
    \leavevmode
    \parbox{0.9\textwidth}
           {\psfig{file=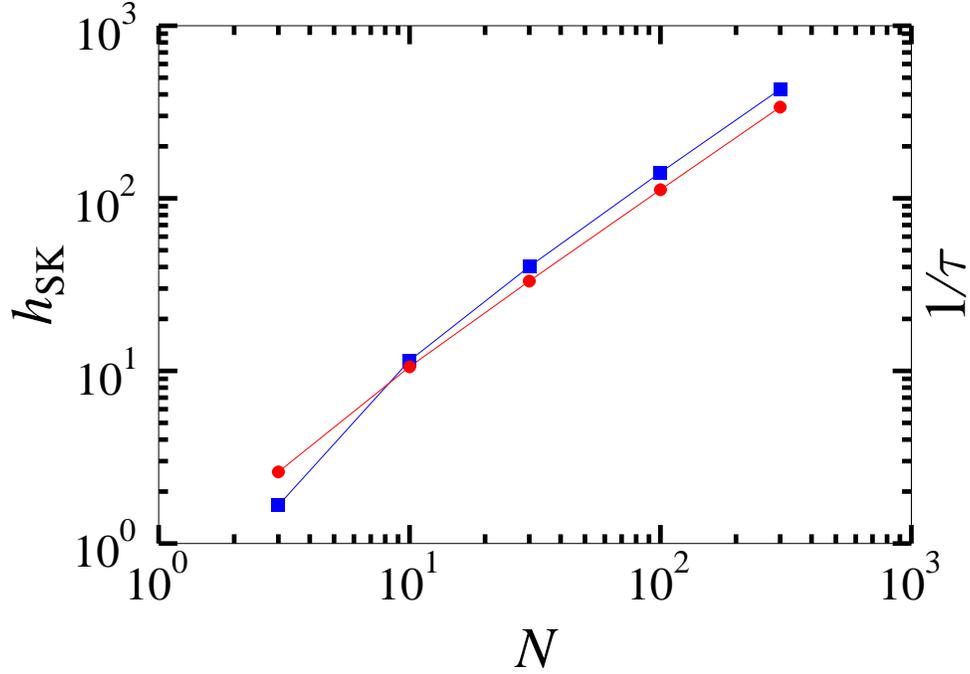,width=0.8\textwidth,angle=0}}
  \end{center}
\protect\caption{KS entropy $h_{\rm SK}$ (squares) and collision rate 
$1/\tau$ (circles) in units of $(E/Nma^2)^{1/2}$ as a function of system 
size $N$.}
\label{fig2}
\end{figure}

\begin{figure}
  \begin{center}
    \leavevmode
    \parbox{0.9\textwidth}
           {\psfig{file=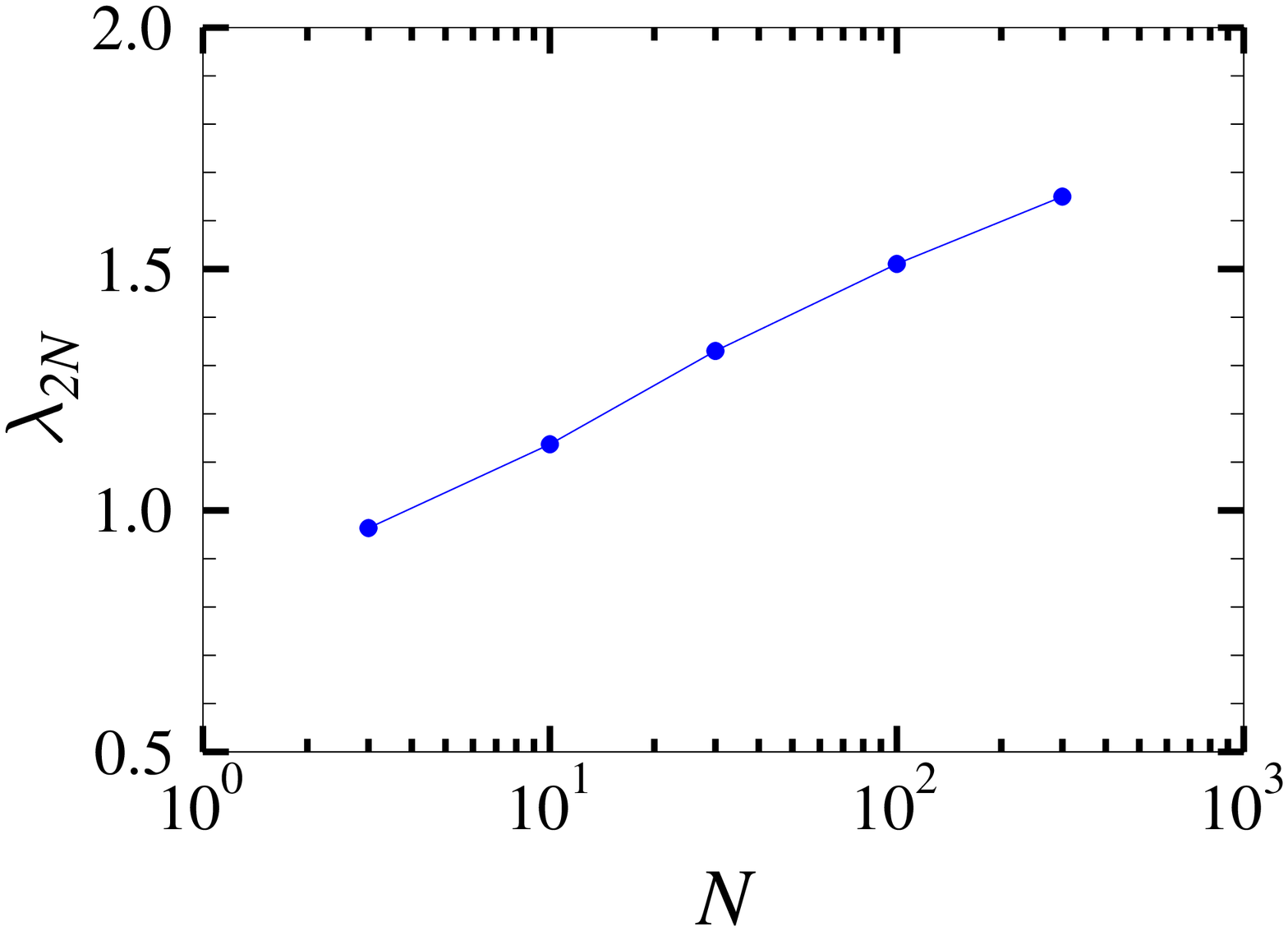,width=0.8\textwidth,angle=0}}
  \end{center}
\protect\caption{Maximal Lyapunov exponent $\lambda_{2N}$ (in units of 
$(E/Nma^2)^{1/2}$) as a function of the system size $N$.}
\label{fig3}
\end{figure}

It is interesting to examine the $N$-dependence in more detail.
Fig.~\ref{fig2} shows that the KS entropy $h_{\rm KS}$ and the collision rate
$\tau^{-1}$ depend linearly on the system size $N$. The case of the collision
rate is easily understood since the constant single particle energy keeps the
collision rate of each particle with the surface constant, too. The KS entropy
is roughly given by the area under the corresponding spectrum presented in
Fig.~\ref{fig1}. Since the spectra converge approximately with increasing $N$
this area increases linearly with the number of particles. The $N$-dependence
of the largest Lyapunov exponent $\lambda_{2N}$ is shown in Fig.~\ref{fig3} and
may be approximated by an logarithmicly increasing curve. In the case of the
hard sphere gas the $N$-dependence could be understood for sufficiently low
densities using techniques borrowed from kinetic theory \cite{vZon}.
Unfortunately, these ideas can not directly be transferred to our system since
the density is not a small parameter. Note however, that the largest Lyapunov
exponent decreases with increasing $N$ once the density is kept constant after
rescaling $a\to a N^{1/3}$. This is interesting with view on nuclear physics
since this result differs qualitatively from simple billiard (mean-field) 
models. Scaling arguments for such models show that the largest Lyapunov 
exponent increases with $N$ at constant density and single-particle energy.

The numerical results obtained in this work indicate that the considered
billiard systems exhibit truly high-dimensional chaos. We recall that the
system is convex and does not possess any dispersing elements. Furthermore, it
differs in construction from the high-dimensional focusing billiards with
spherical caps studied in refs.\cite{RehaBu96,BCG96}. Thus, a closer
examination of the chaos generating defocusing mechanism is of interest and 
presented in the following section.
  
\section{Defocusing mechanism}

Let us examine the defocusing mechanism in the billiard considered in this
work. We do not try to proof that the considered system is completely chaotic
-- which seems difficult at least -- but rather want to understand the
numerically observed phenomenon of chaotic motion in more detail. To this
purpose and based on our numerical results we assume that the system is
(predominantly) chaotic, and that chaos is generated by the only possible
mechanism, namely defocusing \cite{Bunim91}. We may then clarify how
high-dimensional chaos develops and thus understand why we observe $2N-4$
positive Lyapunov exponents. This investigation may hopefully serve also as a
starting point and a motivation for further research.

For simplicity let us consider the three-body system first. It is useful to
study this system as a billiard in full six-dimensional configuration
space. This is possible since the change in relative momentum (\ref{reflect})
caused by an interaction of two particles corresponds to a specular reflection
in the billiard. We denote vectors in configuration space by capital letters as
$\vec{R}=(\vec{r}_1,\vec{r}_2,\vec{r}_3)$, where $\vec{r}_i=(x_i,y_i)$ is the
two-dimensional position vector of the $i^{\rm th}$ particle. The part of the
boundary where particles labelled $i=1, 2$ interact may be parametrized as
\ba
\label{b12}
\vec{X}_{(12)}&=&(\vec{r}+{a\over 2}\vec{e}_\alpha,
                  \vec{r}-{a\over 2}\vec{e}_\alpha,\vec{r}_3), \nonumber\\
\vec{e}_\alpha&=&(\cos{\alpha},\sin{\alpha}).
\ea
The (outwards pointing) normal vector $\partial_a\vec{X}_{(12)}$ and
the tangent vector $\partial_\alpha\vec{X}_{(12)}$ span the
two-dimensional planes where divergence due to defocusing might be generated. 
These planes come in a four-dimensional family due to the parameters $\vec{r}$ 
and $\vec{r}_3$ in eq.~(\ref{b12}). Basis vectors for these planes may be 
chosen as
\ba
\vec{E}_1&=&((1,0),(-1,0),(0,0))/\sqrt{2},\nonumber\\
\vec{E}_2&=&((0,1),(0,-1),(0,0))/\sqrt{2}.
\ea
Similar arguments show that there are two further planes where defocusing
might be generated corresponding to interactions between particles $(1,3)$ and
$(2,3)$, respectively. These planes are spanned by the basis vectors
\ba
\vec{E}_3&=&((1,0),(0,0),(-1,0))/\sqrt{2},\nonumber\\
\vec{E}_4&=&((0,1),(0,0),(0,-1))/\sqrt{2}
\ea
and
\ba
\vec{E}_5&=&((0,0),(1,0),(-1,0))/\sqrt{2},\nonumber\\
\vec{E}_6&=&((0,0),(0,1),(0,-1))/\sqrt{2},
\ea
respectively. Four of the six basis vectors $\vec{E}_i$ are linearly
independent. The vectors $\vec{X}=((1,0),(1,0),(1,0))$ and
$\vec{Y}=((0,1),(0,1),(0,1))$ correspond to displacements of the center of mass
and are orthogonal to the vectors $\vec{E}_i$. This is expected since the
center of mass motion moves freely. It is important to note that the boundary
is neutral (i.e. neither focusing nor dispersing) in the transverse  
directions. 

It is straight forward to generalize these considerations to $N$ bodies. In the
case of the $N$-body billiard there are $N(N-1)/2$ families of two-dimensional
planes where defocusing might possibly occur. These families are related by
those permutations that involve two out of $N$ particles, i.e.
transpositions. $2(N-1)$ out of the $N(N-1)$ basis vectors $\vec{E}_i$ are
linearly independent. The two vectors corresponding to the displacement of the
center of mass are orthogonal to the vectors $\vec{E}_i$.

It would be interesting to relate the number of positive Lyapunov exponents and
the number of linearly independent basis vectors $\vec{E}_i$. Clearly, the
former cannot exceed the latter. Assume that defocusing causes divergence in
the directions of all linearly independent $\vec{E}_i$. Then there would be
exactly $2(N-1)$ positive Lyapunov exponents. However, the conservation of
energy and angular momentum puts two additional constraints, and $2N-4$ is the
number of positive Lyapunov exponents. This reasoning is consistent with the
numerical results presented in the previous section.

The following picture thus arises. The boundary of the billiard considered in
this work consists of several equivalent elements each of which cause a
reflected trajectory to diverge {\it only} in a two-dimensional plane. The
orientation of this plane is determined by the reflecting boundary element.  In
transverse directions the reflection is neutral, i.e. neither focusing nor
dispersing. A trajectory that gets reflected from sufficiently many different
boundary elements may exhibit divergence in all directions. It is interesting
to note that this mechanism differs from the one investigated by
Bunimovich {\it et al.} \cite{RehaBu96,BCG96}. The neutral behavior in the
transverse directions has the advantage that it avoids the problems caused by
the weak convergence occurring in the transversal directions upon reflections
from higher-dimensional spherical caps. It has the disadvantage that several
focusing elements are needed to produce high-dimensional chaos while a single
spherical cap may be sufficient.

\section{Conclusions}

We have computed the Lyapunov spectrum and the KS entropy for an interacting
$N$-body system in two spatial dimensions which is realized as a convex
billiard in $2N$-dimensional configuration space. In presence of four conserved
quantities we find the maximal number of $2N-4$ positive Lyapunov exponents.
Thus, the system exhibits high-dimensional chaos.  At fixed single particle
energy the largest Lyapunov exponent grows with $\ln{N}$ while the KS entropy
grows and the collision rate increase linearly with $N$. In an attempt to
understand the chaotic nature of the billiard we have identified several
symmetry related two-dimensional planes where defocusing might be generated. 
Their number and orientation in configuration space is such that a long
trajectory may exhibit divergence in $2N-4$ directions of phase space.  This
mechanism of focusing differs from the one proposed recently by Bunimovich 
and Rehacek.

Let us finally comment on chaos in realistic many-body systems.  Though the
considered model is a crude approximation of realistic self-bound many-body
systems like nuclei or clusters it incorporates the important ingredient of an
attractive two-body interaction that acts mainly at the surface of the
system. This is the basic picture we have for nuclei and clusters, where the
complicated two-body force creates a rather flat mean-field potential, and
particles experience mainly a surface interaction.  However, unlike in the
model system, an interaction at the nuclear surface involves more than just two
nucleons, and the two-dimensional planes where defocusing is generated in the 
model system are replaced by some higher-dimensional ones. This might also 
introduce the problem caused by weak focusing in transverse directions. It is 
fair to assume that truly high-dimensional chaos may develop upon several 
collisions with the surface. Though the detailed analysis seems much more 
complicated than in the studied model system, the basic picture developed in 
this work should be applicable to some extend also in the case of more 
realistic two-body interactions.

\begin{table}
\begin{tabular}{|r||d|d|d|d|}
$N$   & $\lambda_{5}$ & $\lambda_{2N}$ & $\tau^{-1}$ & $ h_{\rm KS}$ \\\hline
3     & 0.70          & 0.96           &  2.60       &  1.66         \\
10    & 0.22          & 1.14           & 10.6        & 11.5          \\
30    & 0.083         & 1.33           & 33.1        & 40.2          \\
100   & 0.025         & 1.51           & 112.        & 141.          \\
300   & 0.008         & 1.65           & 338.        & 428.          \\ 
\end{tabular}
\protect\caption{Smallest and largest Lyapunov exponents, collision rates and
KS entropy for different system sizes at fixed single particle energy $E/N=1$.
All quantities are given in units of $(E/Nma^2)^{1/2}$.}  
\label{tab1}
\end{table}

\begin{references}  
%
\bibitem{SinaiB}
Ya. G. Sinai,
Sov. Math. Dokl. {\bf 4}, 1818 (1963)
%
\bibitem{BunimS}
L. A. Bunimovich,
Funct. Anal. Appl. {\bf 8}, 254 (1974)
%
\bibitem{RehaBu96}
L. A. Bunimovich and J. Rehacek,
Commun. Math. Phys. {\bf 189}, 729 (1997) 
%
\bibitem{BCG96}
L. Bunimovich, G. Casati, and I. Guarneri,
\prl {\bf 77}, 2941 (1996)
%
\bibitem{GMW}
T. Guhr, A. M\"uller-Groeling, and H. A. Weidenm\"uller,
Phys. Rep. {\bf 299}, 189 (1998)
%
\bibitem{Gaspard}
P. Gaspard,
{\it Chaos, Scattering and Statistical Mechanics},
Cambridge University Press (Cambridge, 1998)
%
\bibitem{DellPoschHoov96}
Ch. Dellago, H. A. Posch, and W. G. Hoover,
\pre {\bf 53} 1485, (1996)
%
\bibitem{vZvBD}
R. van Zon, H. van Beijeren, and J. R. Dorfman,
e-print chao-dyn/9906040
%
\bibitem{TP99}
T. Papenbrock,
e-print chao-dyn/9905007
%
\bibitem{PP99}
T. Papenbrock and T. Prosen,
e-print chao-dyn/9905008
%
\bibitem{Ruelle}
J.-P. Eckmann and D. Ruelle,
Rev. Mod. Phys. {\bf 57}, 617 (1985)
%
\bibitem{Bunim91}
L. A. Bunimovich, 
Chaos {\bf 1}, 187 (1991)
%
\bibitem{BPR99} 
G. F. Bertsch, T. Papenbrock, and S. Reddy,
e-print nucl-th/9906054
%
\bibitem{SWG}
For a review see e.g. 
K. D. Luks and J. J. Kozak,
Adv. Chem. Phys. {\bf 37}, 139 (1978)
%
\bibitem{Prosen}
T. Prosen (private communication)
%
\bibitem{LichtLieb}
A. J. Lichtenberg and M. A. Lieberman,
{\it Regular and Stochastic Motion}, Springer-Verlag, (New York 1983)
%
\bibitem{Reichl}
L. E. Reichl,
{\it The Transition to Chaos in Conservative Classical Systems: Quantum 
Manifestations}, Springer-Verlag, (New York, 1992)  
%
\bibitem{Benettin76}
G. Benettin, L. Galgani, and J. M. Strelcyn,
\pra {\bf 14}, 2338 (1976)
%
\bibitem{Benettin79}
G. Benettin, C. Froeschle, and J. P. Scheidecker,
\pra {\bf 19}, 2454 (1979)
%
\bibitem{Sieber}
M. Sieber,
Nonlinearity {\bf 11}, 1607 (1998)
%
\bibitem{Piesin76}
Ya. G. Piesin,
Math. Dokl. {\bf 17}, 196 (1976)
%
\bibitem{LJ}
H. A. Posch and W. G. Hoover,
\pra {\bf 38}, 473 (1994)
%
\bibitem{FPU}
R. Livi, A. Politi, and S. Ruffo,
J. Phys. A {\bf 19}, 2033 (1986)
%
\bibitem{vZon}
R. van Zon, H. van Beijeren, and Ch. Dellago,
\prl {\bf 80}, 2035 (1998)
%
\end{references}
\end{document}